\documentclass{appolb}
\usepackage[english]{babel}
\usepackage[utf8]{inputenc}
\usepackage{graphicx}
\usepackage{cite}
\usepackage{hyperref}
\usepackage{amsmath}
\usepackage{graphicx}
\usepackage{dcolumn}
\usepackage{bm}
\usepackage{graphicx}

\begin{document}
	
\title{A new and finite family of solutions of hydrodynamics \\
	Part I: Fits to pseudorapidity distributions}
\author{{T. Cs\"org\H{o}$^{1,2}$, G. Kasza$^2$ M. Csan\'ad$^3$ and Z. F. Jiang$^{4,5}$}\\[1ex]
	$^1$Wigner RCP, H - 1525 Budapest 114, P.O.Box 49, Hungary,\\
	$^2$EKU KRC, H-3200 Gy\"ongy\"os, M\'atrai \'ut 36, Hungary,\\
	$^{3}$ELTE, H-1117 Budapest, P\'azm\'any P. s. 1/A, Hungary\\
	$^{4}$Key Lab. of Quark and Lepton Physics,  430079, China\\
	$^{5}$ \quad Inst. Particle Physics, CCNU, Wuhan 430079, China
}	
	
\maketitle
	
\begin{abstract} 
	We highlight some of the interesting properties of a new
	and finite, exact family of solutions of 1 + 1
	dimensional perfect fluid relativistic hydrodynamics.
	After reviewing the main properties of this family of
	solutions, we present the formulas that connect it to
	the measured rapidity and pseudo-rapidity densities and
	illustrate the results with fits to p+p collisions at 8
	TeV and  Pb+Pb collisions at $\sqrt{s_{NN}} = 5.02 $ TeV.
\end{abstract}

\section{Introduction}
In this manuscript we discuss
a new family of exact solutions of perfect fluid 
hydrodynamics for a 1+1 dimensional, longitudinally 
expanding fireball. The applications of 1+1 dimensional
hydrodynamics to particle production in high energy physics
has a long and illustrous history,
that include some of the most renowned theoretical papers 
in high energy heavy ion physics.

In high energy collisions, thermal models to describe
particle production rates were introduced by Fermi in
1950 ~\cite{Fermi:1950jd}.  It was soon pointed out by Landau,
Khalatnikov and Belenkij
~\cite{Landau:1953gs,Khalatnikov:1954,Belenkij:1956cd}, that 
the momentum spectrum can also be explained 
in these collisions  if one
assumes not only global but also local thermal equilibrium.  
Landau and collaborators
predicted~\cite{Belenkij:1956cd},
that perfect fluid hydrodynamical
modelling will be a relevant tool for the analysis of
experimental data of strongly interacting high energy collisions.
After 60 years, this field is still interesting and 
surprizing, as reviewed recently in ref.~\cite{deSouza:2015ena}. 
Applications of exact solutions of relativistic hydrodynamics 
to describe pseudo-rapidity distributions in high energy 
collisions were reviewed recently  in ref.~\cite{Csorgo:2018pxh}.

\section{Equations of relativistic hydrodynamics}
Relativistic perfect fluids are locally thermalized fluids,
their dynamical equations of motion correspond to the
local conservation of the flow of entropy and the flow
of four momentum:
\begin{eqnarray}
	\partial_{\mu}\left(\sigma u^{\mu}\right)&=&0, 
		\label{e:entropy} \\
	\partial_{\nu}T^{\mu \nu} &= &0, 
		\label{e:energy-momentum} 
\end{eqnarray}
where the entropy density is denoted by $\sigma = \sigma(x)$,
four velocity is $u^{\mu}$, normalized as $u^{\mu}u_{\mu}=1$, and the
energy-momentum four tensor is denoted by $T^{\mu \nu}$.  These
fields are functions of  the four coordinate $x^{\mu}=(t,\mathbf{r})
= \left(t,r_x,r_y,r_z\right)$. Similarly, the four momentum is
denoted by $p^{\mu} = (E_p, \mathbf{p}) =
\left(E_p,p_x,p_y,p_z\right)$, where the energy is on mass-shell,
$E_p = \sqrt{m^2 + \mathbf{p}^2}$, where the mass of the  
observed type of particle is indicated by $m$.

The energy-momentum four tensor $T^{\mu\nu}$ of a perfect fluid is
given as  
\begin{equation} 
	T^{\mu \nu}=\left(\varepsilon+p\right)u^{\mu}u^{\nu} - pg^{\mu \nu},
\end{equation} 
where the metric tensor is $g^{\mu
\nu}=\textnormal{diag}(1,-1,-1,-1)$, the energy density  is indicated
by $\varepsilon$ and the pressure by $p$.

The five dynamical equations of relativistic hydrodynamics connect
six variables, the entropy, the energy density, the pressure and the
three spatial components of the four velocity $u^{\mu} = \gamma(1,\mathbf{v})$. This set of equations is closed by the equation of state,
that characterizes the properties of the flowing matter.
We assume, that this is given by
\begin{equation}
        \varepsilon=\kappa p,
\end{equation}
where in this paper,  $\kappa$ 
is assumed to be a constant, independent of the temperature $T$. 
For net baryon free matter, the baryochemical potential is 
$\mu_B = 0$, hence the fundamental thermodynamical relation
reads as $\varepsilon + p = T \sigma$, so the temperature field
can also be chosen as one of the local characteristics of the matter.

In this paper we recapitulate a recent solution
of relativistic hydrodynamics in 
$1$$+$$1$ dimensions,
with a realistic speed of sound 
\begin{equation}
	c_s = 
	\sqrt{\left.\frac{\partial \varepsilon}{\partial p}\right|_{\sigma}}
	\, = \, 1/\sqrt{\kappa} ,
\end{equation}
where in the calculations we use the average value of
the speed of sound, $ c_s = 0.35\pm 0.05$
as measured by the PHENIX Collaboration
in $\sqrt{s_{NN}} = 200$ GeV Au+Au collisions in ref.~\cite{Adare:2006ti}. 

\section{The CKCJ solution}
In $1$$+$$1$ dimensions, it is useful  to rewrite the 
equations of relativistic
hydrodynamics in Rindler coordinates
$(\tau,\eta_x)$ 
~\cite{Csorgo:2006ax,Nagy:2007xn,Csorgo:2008pe,Csorgo:2018pxh}.
The (longitudinal) proper-time $\tau$ and  the coordinate-space
rapidity $\eta_x$ are
\begin{equation}
	\left(\tau,\eta_x\right) =  
		\left(\, \sqrt{t^2-r_z^2}\, ,           
	\frac{1}{2}\textnormal{ln}\left[\frac{t+r_z}{t-r_z}\right]\,\right),
\end{equation}
while the fluid rapidity $\Omega $$=$$ \frac{1}{2}\textnormal{ln}\left(\frac{1+v_z}{1-v_z}\right)$ relates to the four  and to the
three velocity as
$ u^{\mu}$$ =$
$ \left(\cosh\left(\Omega\right),\sinh\left(\Omega\right)\right)$,
$ v_z $$ = $$\tanh\left(\Omega\right)$.

A finite and accelerating, realistic 1+1 dimensional
solution of relativistic hydrodynamics was recently
given by Cs\"org\H{o}, Kasza,
Csan\'ad and Jiang (CKCJ) ~\cite{Csorgo:2018pxh} 
as a family of parametric curves:
\begin{eqnarray}
	\eta_x(H)  & = & \Omega(H) -H, 
		\label{e:etaH}\\ 
	\Omega(H)  & = & 
	\frac{\lambda}{\sqrt{\lambda-1}\sqrt{\kappa-\lambda}}
	\textnormal{arctan}\left(\sqrt{\frac{\kappa-\lambda}
	{\lambda-1}}\textnormal{tanh}\left(H\right)\right), 
		\label{e:OmegaH} \\ 
	\sigma(\tau,H)&= & \sigma_0 
	\left(\frac{\tau_0}{\tau}\right)^{\lambda}
	\mathcal{V}_{\sigma}(s) \left[1+\frac{\kappa-1}{\lambda-1}
	\textnormal{sinh}^2(H)\right]^{-\frac{\lambda}{2}},
		\label{e:sigmasol} \\
	T(\tau,H)  & = & T_0 
	\left(\frac{\tau_0}{\tau}\right)^{\frac{\lambda}{\kappa}} 
	\mathcal{T}(s) 
	\left[1+\frac{\kappa-1}{\lambda-1}\textnormal{sinh}^2(H)
	\right]^{-\frac{\lambda}{2\kappa}},
	\label{e:Tsol}\\ 
	\mathcal{T}(s) & = & 
	\frac{1}{\mathcal{V}_{\sigma}(s)},
	\label{e:scalingsol}\\
	s(\tau,H) & = & 
	\left(\frac{\tau_0}{\tau}\right)^{\lambda-1} 
	\textnormal{sinh}(H)\left[1 + \frac{\kappa-1}{\lambda-1}
	\textnormal{sinh}^2(H)\right]^{-\lambda/2},
		\label{e:sH}
\end{eqnarray}
where the parameter of the solutions, denoted by $H$, stands also
for the difference between the fluid rapidity $\Omega$ 
and the space-time rapidity $\eta_x$. Near mid-rapidity, these solutions
are approximately, but not exactly self-similar~\cite{Csorgo:2018pxh}, they depend
on the space-time rapidity $\eta_x$ predominantly through 
the scaling functions $\mathcal{T}(s)$ and $\mathcal{V}_{\sigma}(s)$,
that in turn depend on the scaling variable $s$.
The solutions  for the fields
$F$$=$$\left\{\sigma, T, \Omega\right\}$, 
and the scaling variable $s$ 
are given with explicit dependence on the longitudinal
proper-time $\tau$ and as parametric solutions in terms 
the parameter $H$. Any of the above
space-time dependent field can be visualized as parametric (hyper)surfaces:
\begin{equation}
	\left(t, r_z, F(t,r_z)\right) = 
	\left(\tau \cosh(\eta_x(H)), \,
	\tau \sinh(\eta_x(H)), \,
	F_s(\tau,H)
	\right), 
\end{equation}
where the subscript $s$ indicates that this function is to be taken
from the  parametric solutions, eqs.~(\ref{e:etaH}-\ref{e:sH}), 
as a function of $\tau$ and $H$.
The functional form of such a bi-variate function $F_s(\tau,H)$,
depends on its variables differently from the functional form of the
also bi-variate function $F(t,r_z)$, as usual.

\begin{figure}
	\includegraphics[scale=0.14]{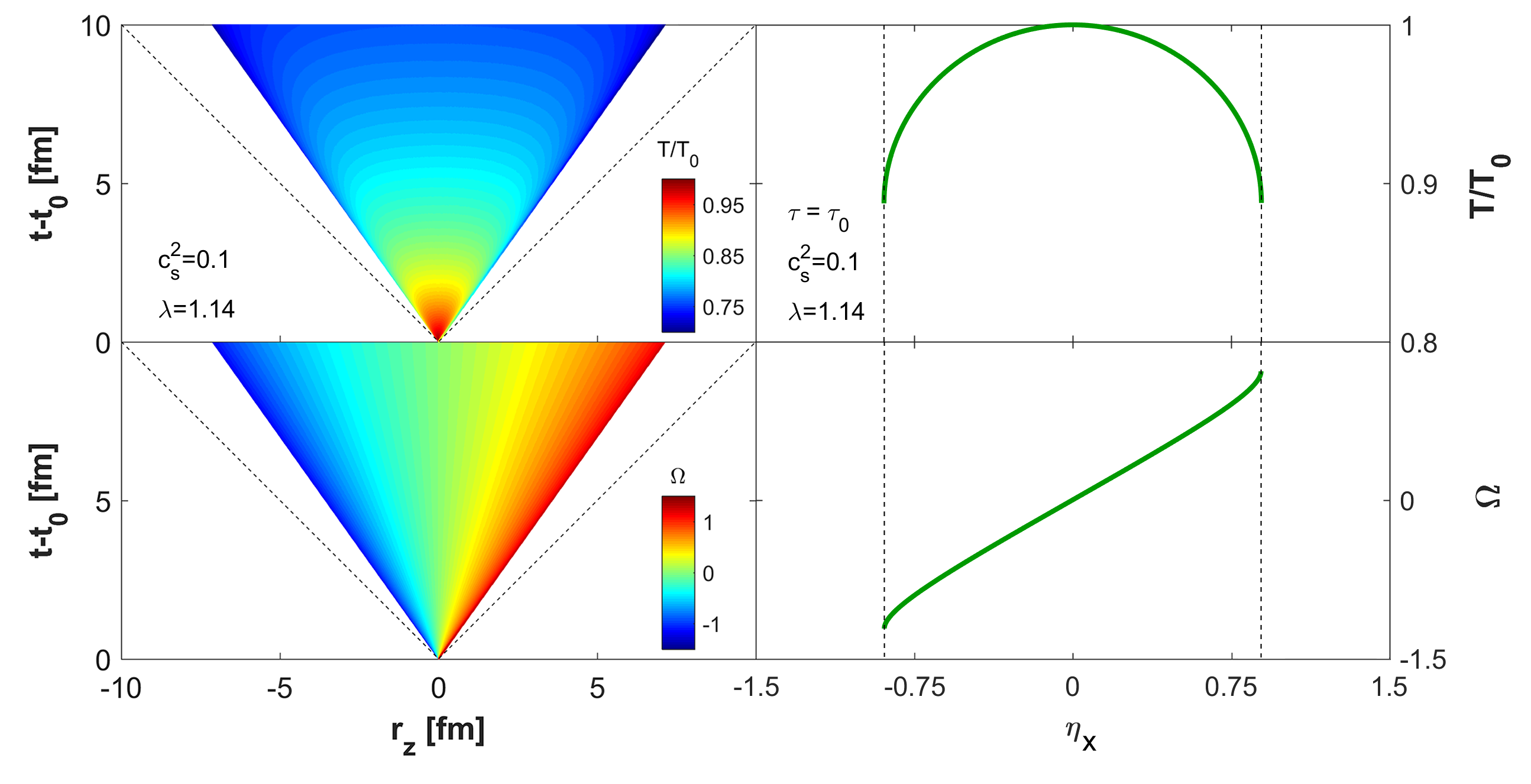}
	\caption{
		Illustration of a CKCJ exact 
		solution~\cite{Csorgo:2018pxh}  of relativistic
		hydrodynamics.  The top left panel shows the
		space-time evolution of the temperature distribution,
		$T(t,r_z)$, while the bottom left panel shows the
		same for the fluid rapidity distribution,
		$\Omega(t,r_z)$.  The top right panel shows the
		temperature at a constant value of the longitudinal
		proper time $\tau$, as a function of the space-time
		rapidity $\eta_x$, where the dashed vertical lines
		indicate the lower and upper limits of the
		applicability of the CKCJ solution. The bottom right
		panel is the same, but it indicates $\Omega(\eta_x)$
		which in this class of solutions is independent of
		the longitudinal proper time $\tau$.
}
	\label{fig:CKCJ_solution}
\end{figure}

This new, longitudinally finite family of solutions is 
illustrated by Fig.~\ref{fig:CKCJ_solution}, for  
a realistic value of the speed of sound,
$c_s^2=1/\kappa = 0.1$ and for a realistic value of
the acceleration parameter, $\lambda=1.14$. 
This figure shows clearly, that the CKCJ solution is limited
to a cone within the forward light-cone around mid-rapidity.
The formulas that give the limiting values of the space-time
rapidity are determined from the 
requirement that the parametric curves of the solution
correspond to functions, as detailed in ref.~\cite{Csorgo:2018pxh}.

\section{Rapidity and pseudo-rapidity distributions}

Let us clarify first the definition of the observables of the
single-particle spectrum in  momentum-space.  The pseudorapidity
$\eta_p$ and the rapidity $y$ of a final state particle with mass $m$
and four momentum $p^\mu$ are defined as
$
\eta_p $$ =$$  \frac{1}{2}\ln\left(\frac{p +p_z}{p-p_z}\right)$ and 
$
y  $$=$$  \frac{1}{2}\textnormal{ln}\left(\frac{E +p_z}{E-p_z}\right),
$
where the modulus of the three-momentum is $p =
|\mathbf{p}| = \sqrt{p_x^2 + p_y^2 + p_z^2}$.

The rapidity and the pseudorapidity distributions were derived from
the CKCJ solutions in ref.~\cite{Csorgo:2018pxh}, as follows. As a
first step, these 1+1 dimensional solutions were embbedded to the 1+3
dimensional space. Subsequently we assumed, that the freeze-out
hypersurface is pseudo-orthogonal to the four velocity and utilized
advanced saddle-point integration methods, to obtain an analytic
expression for the rapidity density distribution~\cite{Csorgo:2018pxh}: 
\begin{equation}
        \frac{dn}{dy} \approx
	\left.\frac{dn}{dy}\right|_{y=0} 
	\cosh^{-\frac{1}{2}\alpha(\kappa)-1}\left(\frac{y}{\alpha(1)}\right)
	\exp\left(-\frac{m}{T_{\rm eff}} 
	\left[\cosh^{\alpha(\kappa)}\left(\frac{y}{\alpha(1)}\right)-1\right]\right),\label{e:dndy}
\end{equation}
where $\alpha(\kappa)$ is defined as $\alpha(\kappa)$$ =$$
\frac{2\lambda-\kappa}{\lambda-\kappa}$.  The mass of the particle
$m$ is the mass of the identified particles (typically pions).  The
above formula depends on four fit parameters, $\kappa$, $\lambda$,
$T_{\rm eff}$  
and  $ \left.\frac{dn}{dy}\right|_{y=0} $. 
These relate to the speed of sound, the acceleration, the effective
temperature (that corresponds to the slope parameter of the invariant
transverse mass spectrum at mid-rapidity), and the value of the
rapidity density at mid-rapidity. 
The values of $T_{\rm eff}$ 
should be determined from fits to the transverse mass spectra of hadrons,
and while $\kappa$  determines the average value of the speed of sound,
measured for example in ref.~\cite{Adare:2006ti}.
The two key parameters of the rapidity density distributions are 
thus the acceleration parameter $\lambda$,
and the midrapidity density, which is just an overall normalization factor. 
Thus the shape of the rapidity distributions is controlled predominantly 
by the acceleration parameter $\lambda$. Both
 $ \left.\frac{dn}{dy}\right|_{y=0} $ and $\lambda$ can be extraced from
fits to experimental data. As the measurement of the
rapidity density distributions requires particle identification, 
the pseudorapidity densities are more readily determined.


Using similar methods, the pseudorapidity density distribution was 
determined as a parametric curve, where the parameter of the curve is
the momentum-space rapidity $y$:
\begin{equation}
        \left(\eta_p(y) \, , \frac{dn}{d\eta_p}(y)  \right) =
        \left( \frac{1}{2}\log\left[\frac{\bar{p}(y) + \bar{p}_z(y)}{\bar{p}(y)-\bar{p}_z(y)}\right] \, , 
         \frac{\bar{p}(y)}{\strut \bar{E}(y)}\frac{dn}{dy} \right),
\label{e:dndeta}
\end{equation}
where $\bar{A}(y)$ denotes  the rapidity dependent average value of
the variable $A$, representing various components of the
four momentum.  The Jacobian connecting the double differential ($y$,
$m_t$)  and ($\eta_p$, $m_t$) distributions has been utilized at the
average value of the transverse momentum, following ref.~\cite{Csorgo:2006ax}. 
In contrast to earlier results, a new element is
that  this CKCJ solution gives an explicit relation between the 
$\bar{p}_T(y)$, the rapidity dependent average transverse momentum,
the slope parameter at mid-rapidity $T_{\rm eff}$  and the mass of the 
observed particles $m$ as follows:
\begin{equation}
	\bar{p}_T (y) \approx 
	{\sqrt{T_{\rm eff}^2 + 2m T_{\rm eff}}}
	\left(1 + 
	\frac{\alpha(\kappa)}{2\alpha(1)^2}
	\frac{T_{\rm eff}+m}{T_{\rm eff}+2m} y^2 \right)^{-1} .
        \label{e:ptbar_y}
\end{equation}
Note, that the same functional form, a Lorentzian shape was obtained 
for the rapidity dependence of the slope of the transverse  momentum
spectrum in the Buda-Lund hydro model of ref.~\cite{Csorgo:1995bi}.
The coefficient of the
$y^2$ dependent term  was considered as a free fit parameter even very
recently, in refs.~\cite{Csanad:2016add,Ze-Fang:2017ppe}. 
This coefficient is now expressed with the help of 
$\kappa$, the parameter of
the equation of state, as well as 
the mass $m$ and the effective slope of the
invariant transverse mass dependent single particle 
spectra $T_{\rm eff}$ at mid-rapidity.  It is remarkable,
that the result of eq. ~(\ref{e:ptbar_y}) is independent of the shape
parameter $\lambda$, that measures the acceleration of the fluid.

\begin{figure}
	\includegraphics[scale=0.44]{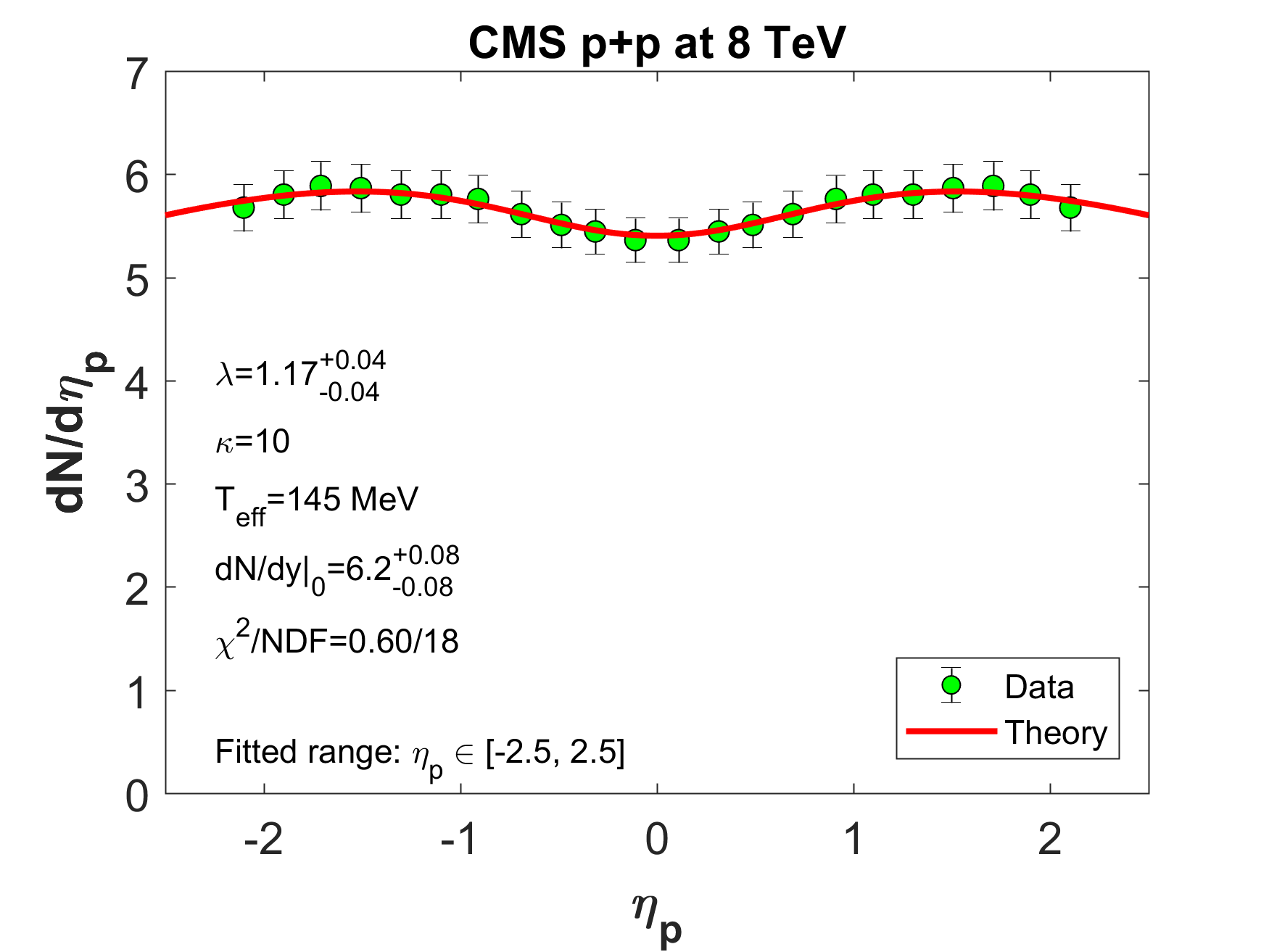}
	\includegraphics[scale=0.44]{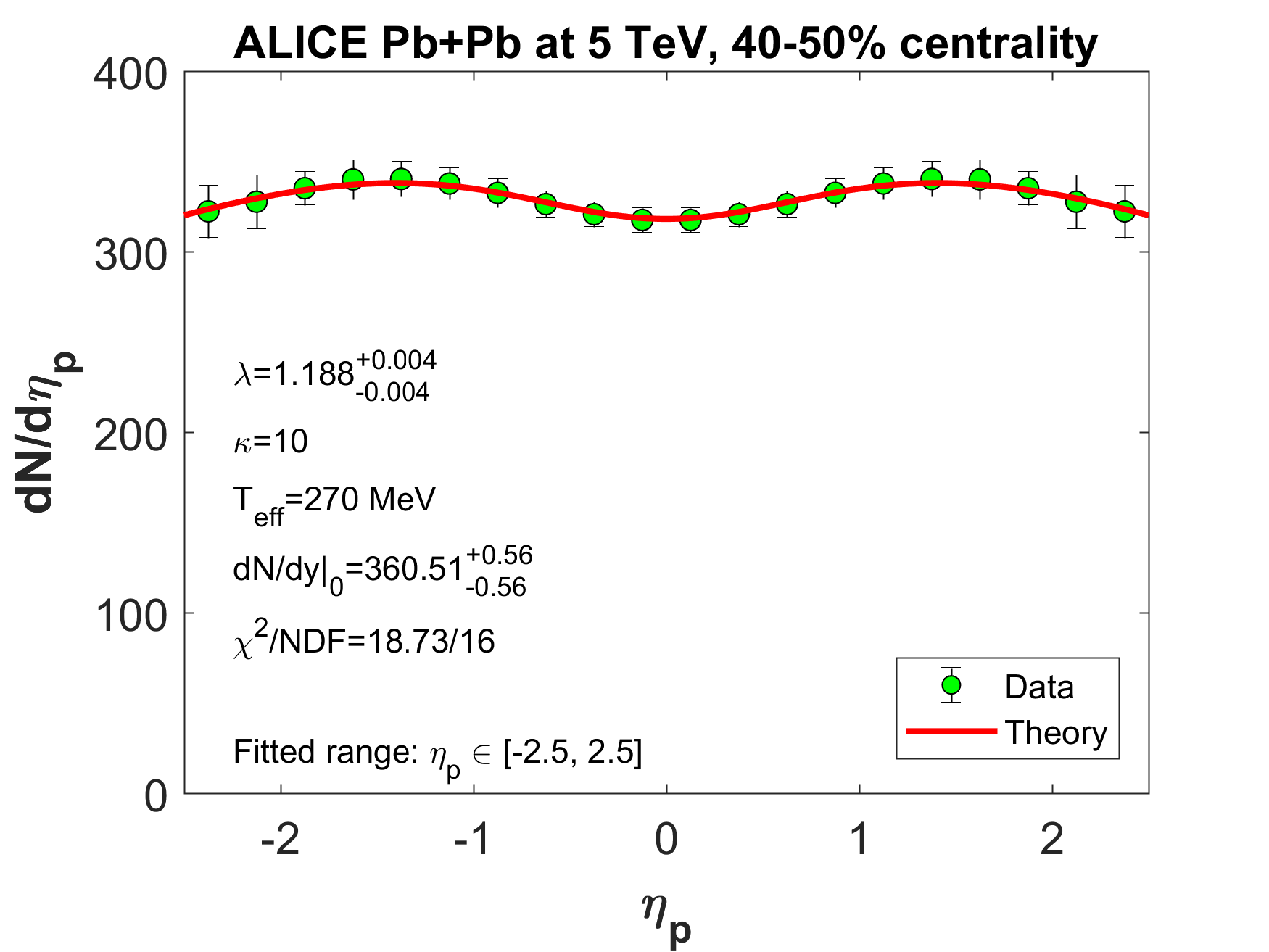}
	\caption{
	(Left) 
	Fits with the CKCJ hydro solution~\cite{Csorgo:2018pxh}, 
	to CMS p+p data at $\sqrt{s} = $ 8 TeV~\cite{Chatrchyan:2014qka}
	using a fixed $T_{\rm eff} = 145 $ MeV.
	(Right) Similar fits, but for ALICE Pb+Pb data at 
	$\sqrt{s_{NN}}$$ =$$ 5.02$ TeV ~\cite{Adam:2015ptt}
	in the 40-50 \% centrality class,
	using  a fixed $T_{\rm eff} = 270$ MeV.
	The speed of sound is $c_s^2 = 1/\kappa = 0.1$, fixed
	in both cases.
	}
	\label{fig:ckcj-fits}
\end{figure}

	The CKCJ hydro solution~\cite{Csorgo:2018pxh} apparently describes the pseudo-rapidity distributions measured by the  
	CMS experiment in p+p collisions at $\sqrt{s} = $ 8 TeV
	~\cite{Chatrchyan:2014qka} in a reasonable manner, 
	for a fixed $T_{\rm eff} = 145 $ MeV,
	as indicated by its fit result on the left panel of  Fig.  ~\ref{fig:ckcj-fits},
	Similarly, the CKCJ hydro solution fits the recent
	ALICE Pb+Pb data at $\sqrt{s_{NN}}$$ =$$ 5.02$ TeV ~\cite{Adam:2015ptt},
	in the 40-50 \% centrality class,
	using  a fixed $T_{\rm eff} = 270$ MeV.
	The speed of sound is fixed in both cases
	to a realistic value of $c_s^2 = 1/\kappa = 0.1$
	~\cite{Adare:2006ti}.

The conditions of validity of these approximations were detailed in 
ref.~\cite{Csorgo:2018pxh}. Typically, these conditions can be simplified for
realistic cases to the condition that the fits are done near to mid-rapidity,
with $|y| < 1/(\lambda -1)$.  For $\lambda $ values reported in this paper,
these conditions are satisfied. Another requirement is that the parametric
curves of these solutions correspond to unique functions of $\eta_x$. Typical
limits from this condition range from $|\eta_x| < 1.0$ to $2.5$. For this reason,
and in order to reduce the effects of fit range dependencies, in this work we
compare fits to various proton-proton and heavy ion collision data by 
limiting the fit range uniformly to $|\eta_x| < 2.5$.

\section{Discussion}
It is interesting to compare the CKCJ solution discussed in the body
of this manuscript to other, well known exact solution of 1+ 1 dimensional solutions of perfect fluid hydrodynamics.

It is rather straight-forward to show, that this class of solutions
includes the Hwa-Bjorken boost-invariant solutions of 
ref.~\cite{Hwa:1974gn,Bjorken:1982qr}, as detailed in
ref.~\cite{Csorgo:2018pxh}. This can be obtained as taking the $H \ll
1$ limiting case first, and subsequently evaluating the $\lambda
\rightarrow 1$ from above limit. In this case, we obtain that
the fluid rapidity $\Omega$ becomes identical with the space-time
rapidity $\eta_x$, the solution becomes boost-invariant and the 
rapidity distribution becomes flat.

It is interesting to note a similarity with Landau's regular
solution, ~\cite{Landau:1953gs,Belenkij:1956cd}
valid also near mid-rapidity, outside the shock-wave region:
In these solutions, the fluid rapidity $\Omega$ and the 
temperature $T$ are used to express the coordinates $(t,r_z)
=(\, t(T,\Omega),\, r_z(T, \Omega))$, while in our CKCJ solutions,
the dependence on the longitudinal proper time $\tau$ is explicitely given, however the dependence on the space-time rapidity $\eta_x$
is given - similarly to Landau's case- as a parametric curve in
terms of the fluid rapidity $\Omega$.

The Cs\"org\H{o}-Grassi-Hama-Kodama (CGHK) 
family of solutions of ref.~\cite{Csorgo:2003rt}
is also recovered easily, in the limit of vanishing acceleration,
that corresponds to $\lambda \rightarrow 1$ from above.

The Cs\"org\H{o}-Nagy-Csan\'ad or CNC family of solutions of 
refs.~\cite{Csorgo:2006ax,Nagy:2007xn}
can be recovered, too, but only carefully, given that in the $\kappa
\rightarrow 1$,  and the $\lambda \rightarrow 1$ limits are 
not interchangeable. First of all, one has to start from
a rewrite of the solutions to the
$1 \le \kappa < \lambda$ domain of the parameters,
which is not discussed here due to space limitations, 
one has to  take the $\kappa \rightarrow 1$ limit
only after this rewrite to recover the CNC solutions.

It is also very interesting to compare our results with the
Bialas-Janik-Peschanski or BJP  solution of 
ref.~\cite{Bialas:2007iu}. A main feature of the BJP solutions
is that the fluid rapidity distribution evolves in time
in an equation of state dependent manner, and approaches
asymptotically the Bjorken limit at every fixed value of the
coordinate $r_z$ for sufficiently late times. In this sense
the BJP solutions initially are similar to a static Landau 
solution (but without the finite lengthscale, the  "l" parameter 
of Landau's solution), while at the end of the time evolution
they asymptotically converge to a Hwa-Bjorken flow velocity field.
Our solutions reviewed here are different in the sense that
as a function of the space-time rapidity $\eta_x$ the fluid rapidity
$\Omega$ is independent of the proper-time $\tau$ so the  
time evolution of the flow field is only apparent, in our case
it is due only to the change of variables from proper-time to time.
A similarity to the BJP solution and to Landau's solution is that
our solution is obtained for an arbitrary but constant value
of the speed of sound.

For more detailed discussions and comparisons of other solutions
with data, we refer to Section 2 of ref.~\cite{Csorgo:2018pxh}.

\section{Summary}

This is the first part of a series of two papers, where
we have highlighted some of the properties of a very recently found,
new family of analytic and accelerating, exact and finite solutions
of relativistic perfect fluid hydrodynamics for 1+1 dimensionally
expanding fireball, evaluated the rapidity and the pseudorapidity
densities from these solutions and demonstrated, that these results
describe well the pseudorapidity densities of proton-proton collisions at 8 TeV colliding energy as measured by the CMS Collaboration
at LHC.  Similarly, this solution also describes the pseudorapidity densities in Pb+Pb collisions at $\sqrt{s_{NN}}$$ = $$5.02$ TeV
measured by the ALICE Collaboration at CERN LHC. These results indicate that the longitudinal expansion dynamics in proton-proton collisions
at CERN LHC is very similar to heavy ion collisions at the nearly the same center of mass energies. 

Our results confirm similar findings, published
recently in ref.~\cite{Ze-Fang:2017ppe},
that was based on the analytically more restricted and simpler, 
1+1 dimensional Csörgő-Nagy-Csanád solutions
of refs.~\cite{Csorgo:2006ax,Nagy:2007xn}
.
These results also suggest
that the space-time rapidity and the fluid rapidity
apparently remain nearly proportional to each other,
even if the speed of sound implemented in two 
different solutions becomes very different from one another. 

\section*{Acknowledgments}
We thank M. Kucharczyk, M. Kłusek-Gawenda and
the LOC of WPCF 2018 for the kind hospitality 
during an inspiring and useful meeting.
M. Csan\'ad was partially supported by the J\'anos Bolyai Research
Scholarship and the \'UNKP-17-4 New National Excellence Program of
the Hungarian Ministry of Human Capacities.
We greatfully acknowledge partial support form the bilateral 
Chinese-Hungarian intergovernmental grant No.~T\'ET 12CN-1-2012-0016,
the CCNU PhD Fund 2016YBZZ100 of China,
the COST Action CA15213 -- THOR Project of the European Union,
the Hungarian NKIFH grants No. FK-123842 and FK-123959,
the Hungarian EFOP 3.6.1-16-2016-00001 project,
the NNSF of China under grant No.~11435004
and the exchange programme of the Hungarian 
and the Ukrain\-ian Academies of Sciences, grants
NKM-82/2016 and NKM-92/2017.

\vfill\eject

\end{document}